\documentstyle[12pt]{article}
\begin{document}
{\
\centerline{\bf Relativistic and QED energy shifts in positronium ion}

\vspace{2cm}
\centerline{ M. Grigorescu and G. W. F. Drake}

\vspace{1cm}
\centerline{ Department of Physics }

\centerline{ University of Windsor}

\centerline{ Windsor, Ontario, Canada }
\vspace{2cm}

{\bf Abstract:} 
The leading relativistic and QED corrections 
to the ground state energy of the three-body system $e^-e^+e^-$ are 
calculated numerically using a Hylleraas correlated basis set. The 
accuracy of the nonrelativistic variational ground state wave function is 
discussed with respect to the convergence properties at the increase 
of the basis dimension and to the variance of the energy expectation value. 
Recent progress in the numerical procedure used to calculate expectation 
values for products of various physical operators is presented. It is shown 
that the nonrelativistic ground state energy can be calculated with an 
accuracy below the level width.  The corrections to this energy include the
lowest order Breit interaction, the vacuum polarization potential, one and
two photon exchange contributions, the annihilation interaction, and
spin-spin contact terms. The relativistic effects and the residual
interactions considered here decrease the one electron binding energy from
the nonrelativistic value of $0.012~005~070~232~980~10(3)$ a.u. to
$0.011~981~051~246(2)$ a.u.. \\[1cm]
{\bf PACS:} 45.50.Jf,36.10.Dr,31.15.-p,12.38.Bx,02.70.-c

\newpage

{\bf I. Introduction} \\[.5cm] \indent
The positronium negative ion (Ps$^-$) is the simplest system composed 
of three equal mass fermions, $e^-e^+e^-$ , bounded only by electromagnetic 
interactions. Similar examples of three-body systems, bounded by 
increasingly complex interactions, are provided by three-quark systems 
such as the proton and neutron, and three-nucleon systems, as the 
$^3$H, $^3$He nuclei.  \\ \indent
In a nonrelativistic approach, accurate numerical approximations to the 
bound eigenstates of three quantum particles interacting by Coulomb forces
can be obtained by using the Rayleigh-Ritz variational method. A suitable
set of coordinates and basis states for the three-body problem was proposed
by Hylleraas \cite{hyll} during the early days of quantum mechanics, and it 
was used to calculate the ground state energy of the helium atom. With 
respect to this set, the matrix elements of various two-body operators 
can be expressed in analytical form \cite{dra1}, and extensive 
high-precision calculations become feasible \cite{dra2} \cite{dra3}.  
 \\ \indent
The relativistic quantm many-body problem can be approached either from the 
field theory, or by using a Schr\"odinger equation with an 
"action at a distance" type Hamiltonian, defined by quantizing the 
classical relativistic system \cite{pamd}. Though, a puzzling result in 
classical mechanics is the no interaction theorem \cite{cjs}, which 
apparently rules out any instantaneous action-at-a-distance Hamiltonian. 
This theorem states that in a classical many-body system, the relativistic 
invariance of the equations of motion (the physical laws) is compatible 
with the "manifest relativistic invariance", of the world lines determined 
by these equations, only if there is no interaction between the particles. 
However, this strong result can be avoided if the interacting particles 
have a structure, as the condition of manifest invariance becomes ambiguous
\cite{hill} \cite{flem}.
 \\ \indent
The approach to the bound state problem based on field theory leads to
a relativistically invariant Bethe-Salpeter equation \cite{beth}, p. 196. In 
the case of two relativistic electrons, approximate Lorentz invariance 
to the first order is introduced by the Breit interaction, which can be 
seen as the quantum correspondent to the Darwin term in classical
electromagnetism \cite{str}. \\ \indent
In the helium atom, the two electrons move in the Coulomb field created by 
a composite, heavy nucleus, which to a first approximation can be 
considered as center of mass (CM). The case of Ps$^-$ is different, because 
all three particles are elementary, have the same mass, and move to
the same degree with respect to the CM. \\ \indent
The existence of a bound ground state in the $e^-e^+e^-$ system was 
predicted by Wheeler \cite{whee} and was observed by Mills \cite{mill} 
passing a positron beam through a thin carbon film in vacuum. The measured 
Ps$^- \rightarrow (2 \gamma) e^-$ decay rate $\lambda =$ 2.09(9) nsec$^{-1}$ 
\cite{mill} corresponds to a Ps$^-$ lifetime of 0.478 nsec, intermediate
between  that of para (singlet) Ps (0.125 nsec) and ortho (triplet) Ps (140 
nsec) \cite{ferr}. \\ \indent
Accurate nonrelativistic numerical calculations for the ground-state
properties of Ps$^-$ are presented in refs. \cite{ho1} to \cite{dra4}.
The autoionization states have been studied in \cite{ho2}, while  
several low-lying resonances have been predicted 
recently \cite{usuk}, by using a combination between the stochastic 
variational method (SVM) with correlated Gaussians and the complex 
scaling method. \\ \indent
The accuracy of the Ps$^-$ groundstate wave functions given by
SVM in a Gaussian basis, was studied by comparison to the direct 
solution of the Schr\"odinger equation in \cite{kriv}. It was shown 
that despite the fact that in SVM the convergence properties of the 
expectation values for most operators are better, the wave function 
is less accurate.
\\ \indent
In this work, the accuracy of the Ps$^-$ nonrelativistic variational
ground state is studied by using beside the convergence properties
of the energy with basis size, also the variance of
the Hamiltonian. The numerical procedure used to calculate matrix elements
is presented in Sect. II. It is shown that in agreement with \cite{kriv},
the variance is larger than the accuracy resulting from convergence. 
Estimates of the relativistic correction terms and the leading QED
corrections are presented in Sect. III. Tables containing the expectation
values of singular operators appearing in the correction terms, such as $p^4$
and delta functions are given in the Appendix. The main results and the
concluding remarks are summarized in Sect. IV. 
\\[.5cm]
{\bf II. The Nonrelativistic Quantum Three-Body Problem}
\\[.5cm] \indent
The nonrelativistic Hamiltonian of the three-body system $e^-e^+e^-$
(or $e^+e^-e^+$) is
\begin{equation} 
H_0 = (- \frac{1}{2} { \nabla}_1^2 - \frac{1}{2} { \nabla}_2^2 -
f { \nabla}_1  \cdot { \nabla}_2  
-\frac{1}{r_1} - \frac{1}{r_2} + \frac{1}{r_{12}} )~ f~ {\rm a.u.}
\end{equation}
where $f= \mu /m$, $ \mu=  m /2$ is the reduced mass, ${ \nabla}_i \equiv
\partial_{ \vec{r}_i}$, $\vec{r}_i=\vec{R}_i/a_\mu$, $i=1,2,3$  denote the
position vectors in the CM frame of the two electrons ($i=1,2$), and of the
positron ($i=3$) in $a_\mu$ units, while $r=\vert \vec{r}_1 - \vec{r}_2
\vert$, $r_1=\vert \vec{r}_1 -\vec{r}_3 \vert$ and $r_2=\vert \vec{r}_2 -
\vec{r}_3 \vert$ are the relative distances. The space coordinates unit is
$a_\mu=a_0/ f$, where $a_0=  \hbar^2 /(m e^2)=0.529~177~249(24)$\AA $~$ is
the Bohr radius. By this choice, the Hamiltonian is naturally expressed in 
reduced atomic units of energy $f$a.u. ($=13.605~698~1(40)$ eV $=1$Ry
if $f=0.5$), where 1 a.u. $= e^2 / a_0  = \alpha^2 mc^2$ is the atomic unit
of energy and $\alpha= e^2 /(  \hbar c )=0.007~297~353~08(32) $
is the fine structure constant.
\\ \indent
Approximate eigenfunctions of this Hamiltonian are obtained by using
the variational method. The trial function is a finite linear combination
\begin{equation}
\Psi(\vec{r}_1, \vec{s}_1; \vec{r}_2, \vec{s}_2) = 
\sum_{a,b,c,l_1,l_2} [
 q_{abc}^{l_1l_2}(1,2)   \Phi_{abc~l_1l_2LM} ( \vec{r}_1, \vec{r}_2)
\end{equation}
$$ 
+ q_{abc}^{l_1l_2}(2,1)   \Phi_{abc~l_1l_2 LM} ( \vec{r}_2, \vec{r}_1) ]
\psi_{Sm_s} ( \vec{s}_1, \vec{s}_2 )
$$
of $N_b$ basis elements $ \Phi_{abc~l_1l_2LM} \psi_{Sm_s}$. The orbital
component $ \Phi_{abc~l_1l_2LM}$ is represented by the Hylleraas
correlated wave function \cite{hyll} 
\begin{equation}
 \Phi_{abc~l_1l_2LM} ( \vec{r}_1, \vec{r}_2) = r_1^a r_2^b r_{12}^c
e^{- \alpha r_1 - \beta r_2} {\cal Y}^{l_1 l_2}_{LM} (1,2) 
\end{equation}
which is a product between a polynomial in all relative radial variables
and the orbital angular momentum eigenstates
\begin{equation}
{\cal Y}^{l_1 l_2}_{LM}(1,2) =
 \sum_{m_1+m_2=M} C^{l_1 l_2 L}_{m_1m_2M}
Y_{l_1 m_1}( \hat{r}_1)  Y_{l_2 m_2}(\hat{r}_2) ~~,
\end{equation}
where $ \hat{r}_i= \vec{r}_i/r_i$, $i=1,2$, $\hat{r}= \vec{r} /r$
are unit vectors. \\ \indent
The spin function
\begin{equation}
\psi_{Sm_s}= \sum_{\mu_1+\mu_2=m_s} C^{ \frac{1}{2} \frac{1}{2} S}_{\mu_1
\mu_2 m_s} \vert \frac{1}{2} \mu_1 \rangle \vert \frac{1}{2} \mu_2 \rangle 
\end{equation}
corresponds to singlet ($S=0$) or triplet ($S=1)$ configurations,
when the orbital part is symmetric ($q_{abc}^{l_1l_2}(1,2)=q_{abc}^{l_1l_2}
(2,1)$) or antisymmetric ($q_{abc}^{l_1l_2}(1,2)=-q_{abc}^{l_1l_2}(2,1)$),
respectively. The expansion coefficients $q_{abc}^{l_1l_2}(1,2)$, and the
non-linear parameters $\alpha$, $\beta$, have been determined previously 
\cite{dra4} by using the variational equations
\begin{equation} 
\delta_{q, \alpha, \beta} \frac{ \langle \Psi \vert H_0 \vert \Psi
\rangle }{
\langle \Psi \vert \Psi \rangle } =0 ~~.
\end{equation}
The kinetic energy operator for the electron 1 is 
$- { \nabla}_1^2/2$, and its effective action in the Hylleraas 
model space is given by
\begin{equation} 
{ \nabla}_1^2 \Phi = [ \frac{1}{r_1^2} \frac{ \partial }{ \partial r_1} 
r_1^2 \frac{ \partial }{ \partial r_1} +  
\frac{1}{r^2} \frac{ \partial }{ \partial r} 
r^2 \frac{ \partial }{ \partial r} - \frac{ \vec{l}_1^2}{r_1^2} 
\end{equation} 
$$
+ \frac{2(r_1-r_2 \hat{r}_1 \cdot \hat{r}_2 )}{r} \frac{ \partial^2 }{ \partial r_1 \partial r}
- \frac{2}{r_1 r} \vec{r}_2 \cdot \nabla_1^Y  \frac{ \partial }{
\partial r}] \Phi 
$$
where $ \vec{l}_1= \vec{r}_1 \times \vec{p}_1$ and $\nabla_1^Y=-i \hat{r}_1 \times \vec{l}_1$. 
A similar expression, obtained by permuting the indices 1 and 2, yields 
${ \nabla}_2^2 \Phi$ . \\ \indent
The action of such operators on the Hylleraas
basis states is complicated, and in the case of their product
becomes tedious, involving hundreds of polynomial terms $r_1^ar_2^br^c$,
spherical harmonics, and singular delta functions. Therefore, in this work 
the matrix elements have been calculated by using a new procedure, 
based on the representation of the physical operators as linear 
combinations within a set ${\cal E}= \{ E_k, k=1,2,...,n \}$ of $n$
elementary operators. Although ${\cal E}$ is not a Lie algebra, this 
approach makes the numerical calculation more flexible, because the 
components $E_k$ can be programmed individually, and they can be assembled 
as needed to form various complicated operators. By a suitable choice, the 
set ${\cal E}$ may account for several physical operators of interest. 
Some  of the elementary operators used in the present calculation are 
presented in Appendix, Table I.  \\ \indent
The accuracy of the wave function improves with the dimension $N_b$ 
of the basis set. When $N_b$ increases, the expectation value of the
Hamiltonian $ \langle H_0 \rangle_{ (N_b) }= \langle \Psi \vert H_0 
\vert \Psi \rangle $ decreases,
and in principle, at the limit $N_b \rightarrow \infty$ the series 
$\langle H_0 \rangle_{(N_b)}$ approaches the exact ground state energy. 
Thus, a measure of the accuracy is provided by the deviation from 0, 
the known limit value, of the "convergence speed"
$[ \langle H_0 \rangle_{(N_b'')}- \langle H_0 \rangle_{(N_b')}]/(N_b''-
N_b')$, where $N_b'$ and $N_b''$ are two consecutive values of $N_b$. A
different measure of the accuracy is the variance 
$\sigma = \sqrt{  \langle H_0^2 \rangle_{(N_b)} - \langle H_0
\rangle^2_{(N_b)} }$. The variational ground state energy $E_g= \langle H_0
\rangle_{(N_b)}$ \cite{dra4} and the present results obtained for
$\sigma^2$ are given in Appendix, Table II. The effective value defined
in the Appendix-(b) is $E_g^{eff}=
-0.524$ 010 140 465 960 215 38$(56)f$ a.u..
Previous estimates of $ E_g $ in Ps$^-$ by using the
correlation-function hyperspherical-harmonic method and the stochastic
variational method are $-0.524~010~139~0~ f$ a.u., respectively
$-0.524~010~140~452~f$ a.u. \cite{kriv}. The effective value obtained in
\cite{frol} by using the extrapolation formula $E_g (N_b) = E_g^{eff} +
A/N_b^p$ is $-0.524$ 010 140 465 956$(8) ~f$ a.u..
\\[.5cm]
{\bf III. Relativistic and QED Corrections}
\\[.5cm] \indent
The quantum description of a relativistic charged fermion based on
Dirac equation requires two spin-1/2 wave functions,
$\vert \psi_+ \rangle$ and $\vert \psi_- \rangle$, corresponding to
the retarded and advanced waves, respectively. For the particle 
eigenstates with energy $E \sim mc^2$, $\vert \psi_- \rangle \sim 
K \vert \psi_+ \rangle$, $K= \vec{ \sigma } \cdot \vec{p} /(2mc)$,          
and the normalization condition
$\langle \psi_+ \vert \psi_+  \rangle + \langle \psi_- \vert 
\psi_- \rangle =1$, can be written in terms of the large components 
as  $\langle \phi \vert \phi \rangle =1$, where $\vert \phi \rangle =
(1 + K^2/2) \vert \psi_+ \rangle$. The antiparticle eigenstates are 
related in principle to the solutions with $E \sim -mc^2$, when $\vert 
\psi_- \rangle$ become the large components. Therefore, in general,
the normalized eigenstate of a nonrelativistic Schr\"odinger equation for 
a particle or antiparticle should be seen as an approximation for $\vert 
\phi \rangle \approx (1 + K^2/2) \vert \psi_L \rangle$, where $ \vert \psi_L 
\rangle $ denotes the corresponding large component. By definition,
if $H_L \vert \psi_L \rangle = E_L \vert \psi_L \rangle$, then
$H_{L,K} \vert \phi \rangle = E_L \vert \phi \rangle$, where 
$H_{L,K}=H_L + [K^2,H_L]/2$ includes both the nonrelativistic 
Hamiltonian and the lowest order correction terms.    
 \\ \indent
In the interacting $e^-e^+e^-$ system $H_{L,K}$ will be restricted to
$H_0 +H_1+H_2$,  where $H_0$ is given by Eq. (1). The term
\begin{equation}
H_1= - \frac{1}{8 m^3 c^2} (p_1^4+p_2^4+p_3^4) 
\end{equation}
($p_i= \vert \vec{p}_i \vert$) 
takes into account the relativistic variation of the mass with velocity,
and
\begin{equation}
H_2=- \frac{ 1}{8 m^2 c^2}  \sum_{i=1}^3   [ \vec{p}_i \cdot , [ \vec{p}_i,
V]]
\end{equation}
derives from the sum between the term $2( \vec{p} V) \cdot \vec{p})/(2mc)^2$
of $H_L$ and the commutator $[K^2,V]/2=2[p^2,V]/(4mc)^2$. 
Here $V(R,R_1,R_2) = e^2 ( 1/R - 1/R_1 - 1/R_2 ) \equiv \tilde{ V} f$ a.u.  and
$R=a_\mu r$, $R_{1,2}=a_\mu r_{1,2}$. \\ \indent
The magnetic current-current interaction plus the retardation
correction corresponding to the lowest-order Breit interaction
are described by the additional term 
\begin{equation}
M_2= - \frac{e^2}{2 m^2c^2} \{ R^{-1}[ \vec{p}_1 \cdot \vec{p}_2 + \hat{r} \cdot (\hat{r} \cdot \vec{p}_1) \vec{p}_2 ]
\end{equation}
$$
- R_1^{-1}[ \vec{p}_1 \cdot \vec{p}_3 +
\hat{r}_1 \cdot (\hat{r}_1 \cdot \vec{p}_1) \vec{p}_3 ]
- R_2^{-1}[ \vec{p}_2 \cdot \vec{p}_3 +
\hat{r}_2 \cdot (\hat{r}_2 \cdot \vec{p}_2) \vec{p}_3 ] \}
$$
such that the total Hamiltonian for Ps$^-$ that will be considered here 
is $H=H_0 +H_1+H_2+M_2$.  \\ \indent
In a classical relativistic many-body system, the dynamical CM defined by the
condition $\sum_i \vec{p}_i =0$, is not necessarily the same as the
geometrical centre of mass, located at $ \vec{R}_{CM} = \sum_i m_i \vec{R}_i
/ \sum_i m_i$. In the present case, if $H=H_0+H_1+H_2+M_2$, only the
dynamical CM is inertial, because $[ \vec{p}_1 + \vec{p}_2 + \vec{p}_3, 
H]=0$, while $[ \vec{r}_1 + \vec{r}_2 + \vec{r}_3, H]$ is not a constant.
The use of the Hylleraas basis ensures that the present calculation 
takes place in the dynamical CM frame, because by the choice of
the coordinates $- i { \nabla}_i = \vec{p}_i /(f \alpha mc)$ and 
${ \nabla}_3 \Phi = (- { \nabla}_1 - { \nabla}_2) \Phi $ for any variational 
wavefunction $\Phi$.  \\ \indent
The expectation value $\langle H_1 \rangle = - (1/64) \langle \nabla_1^4 +
\nabla_2^4 + \nabla_3^4 \rangle  \alpha^2 f$ a.u. can be calculated either
directly, or by assuming that in the ground state 
$ \langle H_0 {\cal O}_p \rangle= \langle {\cal O}_pH_0 \rangle=E_g \langle
{\cal O}_p \rangle$ for any operator ${\cal O}_p$, and using the equalities 
$\nabla_1^2+ \nabla_2^2 = 2( \tilde{H}_0- \tilde{V}+f { \nabla}_1 \cdot 
{ \nabla}_2)$, with $\tilde{H}_0 = H_0/(f {\rm a.u.})$, and
\begin{equation}
\nabla_1^4 + \nabla_2^4=4 ( \tilde{H}_0- \tilde{V} + 
f { \nabla}_1 \cdot { \nabla}_2)^2 
-2 \nabla_1^2 \nabla_2^2 ~~,
\end{equation}
\begin{equation}
\nabla_3^4=(- \vec{ \nabla}_1- \vec{ \nabla}_2)^4=\nabla_1^4+ \nabla_2^4+
4( { \nabla}_1 \cdot { \nabla}_2)^2 
\end{equation}
$$
+ 2 \nabla_1^2 \nabla_2^2 + 4 (\nabla_1^2+ \nabla_2^2) 
{ \nabla}_1 \cdot { \nabla}_2 ~~.
$$ 
Although formally the same, within a finite basis the two expressions give 
slightly different results ($ \langle \nabla_1^4 \rangle, \langle \nabla^4_1
\rangle_E $), presented in the Appendix, Table III. In the numerical
estimates we have used only $ \langle \nabla^4_1 \rangle_E $, because of
its rapid convergence and higher acuracy in the effective value.
\\ \indent
The term $H_2 $ contains the singular operators
$ \Delta_1 \tilde{ V }= - 4 \pi [ \delta( \vec{r} ) -
\delta( \vec{r}_1)] $,
$ \Delta_2 \tilde{ V } = - 4 \pi [ \delta( \vec{r} ) -
\delta( \vec{r}_2)] $, and
$ \Delta_3 \tilde{ V } =  4 \pi  [ \delta( \vec{r}_1 ) +
\delta( \vec{r}_2)] $,
which yield
\begin{equation}
\langle H_2 \rangle = \alpha^2 \pi \langle \delta( \vec{r}_1 )+ \delta( \vec{r}_2 )
- \delta( \vec{r} ) \rangle f^3 ~~ {\rm a.u.}
\end{equation}
Previous estimates of $\langle \delta ( \vec{R}_1) \rangle$ ($= a_\mu^{-3}
\langle \delta ( \vec{r}_1) \rangle $) in Ps$^-$ by using the
correlation - function hyperspherical - harmonic method and the stochastic
variational method are 0.020 733 14$(6) a_0^{-3}$,
respectively
0.020 731 048 976 $a_0^{-3}$ \cite{kriv}. The same methods give for
$\langle \delta ( \vec{R} ) \rangle $ the values 0.000 170 997$(2) a_0^{-3}$
and  0.000 171 112 600 741 $ a_0^{-3}$, respectively \cite{kriv}.
The results of the present calculation, in the same units ($a_0^{-3}$), 
are listed in Appendix, Table IV as a function of the dimension $N_b$ of the
basis set. \\ \indent  
The expectation values which appear in the calculus of $\langle M_2 \rangle$,
obtained when $N_b=324$ are  
$$
u_{ee}= \langle r^{-1} { \nabla }_1 \cdot { \nabla}_2 \rangle = -0.008~267~646~67 ~~,
$$
$$
v_{ee}= \langle r^{-1} \hat{r} \cdot (\hat{r} \cdot { \nabla}_1)
{ \nabla}_2 \rangle = 0.019~610~925~35
$$ 
and for $i=1,2$  
$$u_{ep}= \langle r^{-1}_i ( { \nabla }_i \cdot { \nabla}_3
\rangle = 1.535~434~049~31
$$
$$
v_{ep}= \langle r^{-1}_i \hat{r}_i \cdot (\hat{r}_i \cdot { \nabla}_3)
 { \nabla}_i \rangle =-0.555~009~821~912 ~~.
$$ 
In terms of these variables, $\langle M_2 \rangle = 0.5 \alpha^2 w f^3 $ a.u. with
$w=u_{ee}+v_{ee}-2u_{ep}-2v_{ep}=-1.949~505~176~125$. For the 2528-dimensional
basis set $w=-1.949~505~250~368$. The average of the last three consecutive
values, obtained for $N_b=1990,2276$ and 2528 gives the effective matrix
element
$w^{eff} =-1.949~505~250~368(1) $. \\ \indent
The effective sum of the spin independent relativistic correction terms
$\langle H_1 \rangle$, $\langle H_2 \rangle$ and $\langle M_2 \rangle$ 
is $ -0.145~476~184~397(8)  \alpha^2 f$
a.u.  which decrease the Ps$^-$ ground state energy to
\begin{equation}
E_{g^*}^{eff} = \langle H \rangle^{eff} =E_g^{eff} - 0.145~476~184~397(8) 
\alpha^2 f \rm{a.u.}~.
\end{equation}
\indent
The same calculations yield for the corrected ground state energy
$E_{g^*}^0$ in neutral positronium $E_{g^*}^0= -(0.5+ 5 \alpha^2/32 ) f$ a.u..
However, for this relativistic two-body system the finite mass 
corrections to the energy provided by the one-body Dirac equation can be 
obtained exactly up to the order $\alpha^2$Ry by using the formula 
\cite{sapi}
\begin{equation}
E_{(n,j,Z)}= \frac{1}{ \alpha^2} [ \eta -1 - \frac{f}{4} ( \eta-1)^2]~~
f~{\rm a.u.}
\end{equation}
where $\eta=1/ \sqrt{ 1+ (Z \alpha)^2 /( n- \nu)^2 }$,  
$ \nu = j+1/2- \sqrt{ (j+1/2)^2-(Z \alpha)^2 } $. The expansion
$E_{(1,1/2,1)} \approx  -(0.5 + 5 \alpha^2 /32)
f$ a.u.  reproduces $E_{g^*}^0$, showing that the relativistic corrections
given by the expectation value of $H$ are reliable.
\\ \indent
Within QED the constituents of the three body-system $e^-e^+e^-$ cease
to be "elementary", because they are subject not only to the mutual two-body 
Coulomb-Breit interaction, but are also coupled to the vacuum fluctuations 
of the electromagnetic field $\vec{A}$ \cite{bjdr}. The interaction terms accounting 
for this coupling are represented by an infinite series of increasingly 
complicated Feynman diagrams with closed photon lines. However, the 
complexity is increased recursively, by taking into account at each order 
three basic processes, represented by the anomalous magnetic moment (vertex)
 corrections, electron self-mass and vacuum polarization diagrams.
 \\ \indent
Although formally complicated, the main effect of the coupling to the field 
degrees of freedom is simply a change in the charge and mass parameters $e$ 
and $m$ of the theory. This contribution has already been taken into account, 
because it is included in the measured values of $e$ and $m$ used to define 
the atomic unit of energy. Though, the QED corrections in the 
interacting three body-system $e^-e^+e^-$ are not the same as for the free
particles, and the differences still need to be considered. \\ \indent
The vacuum polarization properties have been studied first by
Heisenberg \cite{heis} and Uehling \cite{uehl}, showing that a given 
charge density $\rho( \vec{R})$ induces a polarization charge 
$\delta \rho( \vec{R}) = - ( \alpha / 15 \pi) \lambda_0^2 { 
\nabla}^2 \rho ( \vec{R} ) $, where $\lambda_0 = \hbar/ mc$ is the Compton
wavelength of the electron. The induced charge leads to deviations from the
standard Coulomb interaction. Thus, the vacuum behaves as an inhomogeneous
dielectric, in which the mutual potential energy between two point-like
charges $Z_1$ and $Z_2$ is \cite{uehl}
\begin{equation}
V(R) = \frac{Z_1 Z_2 e^2}{R} [ 1 - \frac{ \alpha}{ \pi} R U(R)]~~,
\end{equation}
by $U(R)$ denoting the Uehling potential. This potential is singular at $R=0$,
falls of exponentially for $R>0$, and satisfies the integral condition 
$\int d^3R U(R) = - 4 \pi \lambda_0^2 /15$. Therefore, it can be well 
approximated by a delta function, $U(R) = - 4 \pi ( \lambda_0^2/ 15) 
\delta( \vec{R})$. In the case of Ps$^-$, the correction term introduced by 
this potential is
\begin{equation}
\langle H_{vp} \rangle = \frac{4}{15} \alpha^3  \langle \delta ( \vec{r} ) - 
\delta( \vec{ r}_1 ) - \delta( \vec{r}_2 )  \rangle f^3 ~{\rm a.u.}
\end{equation}
By using the effective values given in Appendix, Table IV,
the contribution of the vacuum polarization to the Ps$^-$ ground
state energy is  $\langle H_{vp} \rangle^{eff} = - 0.022~024~212~934~6(7)
\alpha^3 ~f$ a.u..
It is important to remark that this value takes into account the positron
recoil (the "mass polarization" term) because the wave functions are 
obtained by minimizing the full nonrelativistic Hamiltonian.    
In neutral positronium, $\langle \delta ( \vec{R}_1) \rangle_{Ps}= 1/ (
\pi a_\mu^3)= 1/(8 \pi a_0^3)$, and the vacuum polarization
correction is $ - 1/(15 \pi) ~ \alpha^3 ~f $ a.u.. 
\\ \indent
As it was shown early by the Lamb shift measurements \cite{lamb}, the main
QED correction appears however from the coupling to the vacuum fluctuations
of the field rather than from the vacuum polarization ( \cite{bjdr} p. 59). 
For a free electron the ground state energy is given by its 
rest mass $m = m_b + \delta m$, consisting of the uncoupled value $m_b$ 
and the positive renormalization term $\delta m = (3 \alpha m_b / 2 \pi) 
\ln ( \Lambda /m_b)$ due to the electromagnetic self-energy, where
$\Lambda$ is a large (formally infinite) cutoff mass. \\ \indent
Similarly, the coupling to the field modes also affects the intrinsic
excitations of a many-body system.
In a bounded $N$-particle system, the shift $\Delta E_n$ in the 
energy $E_n= \langle n \vert H_0 \vert n \rangle$ of the level $\vert n
\rangle$ due to the exchange of a transverse photon can be obtained by
using the time-independent second-order perturbation expression 
\begin{equation}
\Delta E_n =-  \langle n, 0_f \vert H_c( \vec{A} ) \frac{1}{H_0 + H_A -E_n} 
H_c( \vec{A}) \vert n, 0_f \rangle ~~.
\end{equation}
Here  $H_c( \vec{A}) = \sum_{i=1}^N h_i( \vec{A} )$ is the sum over all
particles of the one-body coupling terms $h_i( \vec{A} )= -e_i \vec{
\alpha}_i \cdot \vec{A}_{( \vec{R}_i)} $, $\vec{ \alpha}_i \approx 
\vec{p}_i/(m_ic)$,
\begin{equation}
\vec{A}_{( \vec{r})} = \frac{ \sqrt{ \hbar c} }{2 \pi}
 \int  \frac{ d^3 k}{\sqrt{k}} \sum_{\lambda=1,2} 
\vec{ \epsilon}_\lambda ( 
\hat{a}^{ \dagger}_{ k \lambda} e^{ - i \vec{ k} \cdot \vec{r} } +   
\hat{a}_{ k \lambda} e^{ i \vec{ k} \cdot \vec{r} } )   
\end{equation}
is the quantized transverse vector potential of the photon
($\vec{\epsilon}_\lambda \cdot \vec{k} =0$, $\vec{ \epsilon}_\lambda^2=1$),
$H_A=\int d^3 k \sum_{\lambda=1,2} \hbar c k a^{\dagger}_{k \lambda}
a_{k \lambda} $ is the free field Hamiltonian, and $\vert 0_f \rangle$
denotes the photon vacuum.  This shift has the form $\Delta E_n =
\sum_{i=1}^N X^n_i + \sum_{i < j} Y^n_{ij}$, where  
\begin{equation}
X^n_i =- \langle n, 0_f \vert h_i( \vec{A} ) \frac{1}{H_0 + H_A -E_n} 
h_i( \vec{A}) \vert n,0_f \rangle
\end{equation}
and
\begin{equation}
Y^n_{ij} = -2 {\rm Re}[ \langle n,0_f \vert h_i( \vec{A} ) \frac{1}{H_0 + H_A -E_n}
h_j( \vec{A})  \vert n,0_f \rangle] ~~.
\end{equation}
\indent
It is important to remark that the interaction with the vacuum field
fluctuations may affect not only the intrinsic dynamics, but also the center
of mass. In a classical two-body system coupled to the field, $H_c$
can be written in terms of the canonical pairs
$( \vec{r}, \vec{p}_\mu ) \equiv 
( \vec{R}_1 - \vec{R}_2,  \mu  \vec{p}_1 /m_1 - \mu \vec{p}_2 /m_2)
$
and
$
( \vec{R}_{CM}, \vec{P}_{CM} ) \equiv ( \mu  \vec{R}_1 /m_2- \mu \vec{R}_2/m_1,
\vec{p}_1+ \vec{p}_2 )
$
of intrinsic and respectively,  center of mass variables, as
$$
\vec{p}_\mu \cdot [ \frac{e_2}{m_2} \vec{A}_{(R_2)}
- \frac{e_1}{m_1} \vec{A}_{(R_1)}
 ] - \frac{ \mu}{m_1m_2} \vec{P}_{CM} \cdot
 [ e_1 \vec{A}_{(R_1)} + e_2 \vec{A}_{(R_2)} ]~~.
$$
This expression shows that in a neutral two-body system (such as Ps) the
center of mass energy is not affected by the field 
only if $ \vec{A}_{(R_1)} = \vec{A}_{(R_2)} $, or when
the size of the system is negligible compared to the photon wavelength
(dipole approximation).   \\ \indent
In a quantum $N$-body system it is convenient to take advantage of the
finite size effects by  writing $\vec{A}$ as the incoherent
sum of long and short wavelength components, $\vec{A}_L$ and $\vec{A}_S$,
obtained by decomposing  $\int d^3k$ as $ \int_{\vert \vec{k} \vert
\le k_L} d^3k + \int_{k_L< \vert \vec{k} \vert <k_M}$,
where $k_L$ and $k_M$ are cutoff parameters. Each domain
brings its own contribution to the matrix elements, which can be
similarly decomposed  as
\begin{equation}
X^n_i= X^{Ln}_i+ X^{Sn}_i ~~,~~ Y^n_{ij}= Y^{Ln}_{ij}+ Y^{Sn}_{ij}~~.
\end{equation}
At the end of the
calculation $k_L$ should disappear, while $k_M \rightarrow \infty$. \\ \indent
If $H_0$ consists of the kinetic energy term plus a local potential $V$,
then a non-relativistic calculation within the dipole approximation yields 
\begin{equation}
X^{Ln}_i =- \frac{ \alpha }{ 3 \pi m^2c^2} [  2 \hbar c k_L \langle n \vert
 \vec{p}_i^2 \vert n \rangle + \langle n \vert  
[ \vec{p}_i \cdot, [  \vec{p}_i, V]] \vert n \rangle 
\ln \frac{ k_L }{k_R} - 2 B^n_{ii}] ~~,
\end{equation}
where $ k_R = {\rm R_M} / \hbar c$, ${\rm R_M}$ is a dimensional constant
with units of energy, and $B^n_{ii}$ are the diagonal elements of the matrix
$[B^n_{ij}]$ defined by
\begin{equation}
B^n_{ij}= \sum_m (E_n-E_m) {\rm Re} ( \langle n \vert \vec{p}_i \vert m \rangle
\cdot \langle m \vert \vec{p}_j \vert n \rangle ) \ln \frac{ \vert E_n-
E_m \vert}{ {\rm R_M}}~~.
\end{equation}
The first term depends only on the kinetic energy, and it can be written
as $- \delta m_L \langle n \vert  \vec{p}_i^2 \vert n \rangle /( 2 m^2 )$,
$\delta m_L= 4 m r_e k_L / (3 \pi ) $, where $r_e= \alpha \hbar /(mc)$
denotes the classical radius of the electron. 
It contributes also to the energy of a free particle ($V=0$) and has the 
structure of a first-order perturbation shift induced by a variation 
$\delta m_L$ of the nonrelativistic mass. Thus, such terms can be taken into
account simply by a redefinition of the cutoff mass $\Lambda$.
\\ \indent
A relativistic calculation of the one-body QED correction arising from the 
exchange of a transverse hard photon at a Coulomb vertex 
\cite{fuma}, \cite{bjdr} p.177, yields 
\begin{equation}
X^{Sn}_i=
\frac{  \alpha \hbar^2}{ 3 \pi m^2c^2} ( \ln \frac{mc}{2 \hbar k_L}
+ \frac{5}{6} ) \langle n \vert \Delta_i V \vert n \rangle ~~,  
\end{equation}
($5/6=11/24+3/8$) such that
\begin{equation}
X^n_i = - \frac{ \delta m_L}{ 2 m^2}
\langle n \vert  \vec{p}_i^2 \vert n \rangle
+
\frac{ \alpha  }{ 3 \pi m^2c^2}
[  \hbar^2 ( \ln \frac{mc}{2 \hbar k_R} + \frac{5}{6})  \langle
\Delta_i V \rangle_n + 2B^n_{ii} ]   ~~.
\end{equation}
The quantity $Y^{Ln}_{ij}$ can be expressed as
\begin{equation}
Y^{Ln}_{ij}=- \frac{  \delta m_L }{ m^2 } \frac{ e_ie_j}{e^2} \langle 
\vec{p}_i \cdot \vec{p}_j \rangle_n -
\frac{ 2 \alpha }{ 3 \pi m^2c^2} \frac{ e_ie_j}{e^2}  
\{ \langle [ \vec{p}_i \cdot, [  \vec{p}_j, V]] \rangle_n
\ln \frac{ k_L }{k_R}  -2 B^n_{ij} \} ~.
\end{equation}
\\ \indent
In the case of Ps$^-$ there are three terms $X^n_i$, one for each
electron ($i=1,2$) and one for the positron ($i=3$), and three terms
$Y^{Ln}_{ij}$, $i<j$. The contribution to
$ \Delta E_n$ arising from the terms linear in $ \delta m_L$ of $X$ and $Y^L$
is $ \delta_0 E_n= - \delta m_L \langle n \vert
(\vec{p}_1 + \vec{p}_2 - \vec{p}_3 )^2  \vert n \rangle /( 2 m^2)$.
In the dynamical CM frame, this energy shift can be accounted, for example,
by an effective variation $3 \delta m_L $ in the total mass of the
electron-electron pair and $ \delta m_L$ in the mass of the positron, or by
a variation of $4 \delta m_L$ in only one of them.
\\ \indent
The definition of the Bethe logarithm $\beta_n \equiv
2 B^n_{33}/ \langle [ \vec{p}_3, \cdot [ \vec{p}_3 , H_0]] \rangle_n$,
and the identity $m^2 \sum_{i,j} e_ie_j B^n(i,j)/(m_i m_j) = e^2 B^n(3,3) 
(1+m/m_3)^2 $ (valid if $m_1=m_2=m$ and $\langle n \vert \sum_i \vec{p}_i  
\vert n' \rangle=0$ for any $n,n'$), show that $\sum_i X^{n}_i + \sum_{i<j}
Y^{Ln}_{ij} = \delta_0 E_n + \delta_1 E_n +  \delta_{2L} E_n $, where
\begin{equation}
 \delta_1 E_n =  \frac{ \alpha \hbar^2 }{ 3 \pi m^2c^2}
 [- 4 \beta_n \langle \Delta_3 V \rangle_n +
( \ln \frac{mc}{2 \hbar k_R} + \frac{5}{6})
\sum_{i=1}^3 \langle \Delta_i V \rangle_n ]  
\end{equation}
$$
 =  \frac{ 4 \alpha^3 }{ 3 }
 [- 4 \beta_n \langle \delta^3(r_1) + \delta^3(r_2) \rangle_n 
+ 2 ( \ln \frac{mc}{2 \hbar k_R} + \frac{5}{6})
\langle \delta^3(r_1)  + \delta^3(r_2) - \delta^3(r) \rangle_n]f^3~{\rm a.u.}
$$
and $ \delta_{2L} E_n \equiv \sum_{i<j} \langle W^L_{ij} \rangle_n$ is
given by the expectation value of the potential
\begin{equation}
W^L_{ij}(k_L) =   \frac{ 8 \alpha^3 }{ 3}  
 \ln \frac{k_L}{k_R} \delta^3( r_{ij} ) ~ f^3~{\rm a.u.}.
\end{equation}
\indent
The term $Y^{Sn}_{ij}$ due to the exchange of a short wavelength
(hard) transverse photon between different particles will be decomposed as
$Y^{Sn}_{ij}=Y^{S2n}_{ij}+Y^{S3n}_{ij}$,  according to the expansion
$1/(H_0+H_A-E_n) \approx 1/H_A - (H_0-E_n)/(H_A)^2$. 
The contribution from $1/H_A$ is
\begin{equation}
Y^{S2n}_{ij} = 
- 2 \langle n,0_f \vert 
h_i( \vec{A}_S ) H_A^{-1} h_j( \vec{A}_S)  \vert n,0_f \rangle ~~.
\end{equation} 
In the limit $k_L \rightarrow 0$, $k_M \rightarrow \infty $, the integral 
over $k$ in this matrix element can be evaluated by using the identity
$$
\int  \frac{ d^3 k}{k^2} e^{i \vec{k} \cdot \vec{r}} ( \vec{A} \cdot \vec{B}
- \hat{k} \cdot \vec{A} \hat{k} \cdot \vec{B} ) = \frac{ \pi}{r}
( \vec{A} \cdot \vec{B} + \hat{r} \cdot \vec{A} \hat{r} \cdot \vec{B})~~,
$$
showing that the sum $\sum_{i < j} Y^{S2n}_{ij}$ becomes the two-body term
$\langle M_2 \rangle_n$ of order $\alpha^2$Ry, already taken into account.
Thus, the only new contribution is the next-order term
\begin{equation}
Y^{S3n}_{ij}= 2 {\rm Re}[ \langle n, 0_f \vert
h_i( \vec{A}_S ) \frac{H_0-E_n}{H_A^2} h_j( \vec{A}_S)
\vert n,0_f \rangle ] 
\end{equation} 
which is the expectation value of the two-body potential
\begin{equation}
W^S_{ij}(k_L,k_M) = 
  \frac{2 \alpha^3 }{3 \pi } [ \frac{3}{2} f(k_L,k_M,r_{ij}) 
+ 4 \pi \delta^3(r_{ij}) \ln \frac{k_M}{k_L} ] f^3 ~a.u.
\end{equation} 
Here $f(k_L,k_M,r) = 2[j_0(k_Mr)+j_2(k_Mr) -j_0(k_L r)-j_2(k_L r)]/(3r^3)$
is the function introduced by Araki  \cite{ara}, written in terms of
the spherical Bessel functions $j_0,j_2$. 
When $k_L \rightarrow 0$, $k_M \rightarrow \infty $, 
$f(0, \infty, r) =  - 2/(3r^3 ) $, but the logarithmic
factor in the second term of $W^S_{ij}$ is divergent at both limits. 
However, the divergence in $k_L$ is cancelled by the low-energy term,
and the sum $W_{ij}(k_M)=W^L_{ij}(k_L)+W^S_{ij}(k_L,k_M)$,
\begin{equation}
W_{ij}(k_M)=
\frac{2 \alpha^3 }{3 \pi } [ 
\frac{3}{2} f(0,k_M,r_{ij}) 
+ \frac{ \delta(r_{ij}) }{r_{ij}^2} \ln \frac{k_M}{k_R} ] f^3 ~a.u.~,
\end{equation}
is independent of $k_L$. The divergent 
factor containing $k_M$ contributes only when $\vert n \rangle$ is an 
$S$ state, but in this case the expectation value $\langle 1/r^3 \rangle_n$ 
is also logarithmically divergent. It is however possible to define a limit
for the sum of these infinite terms in the sense of the principal value. Let 
\begin{equation}
D(a,r) = \frac{ \theta(r-a)}{r^3} - \frac{ \delta(r)}{r^2} \ln
\frac{a_\mu}{a}
\end{equation}
be a distribution depending on the positive radius parameter $a=
\eta /k_M$, where $\eta$ is a positive scale factor.
Because $r^2 \partial_a D(a,r) = [ \delta(r)- \delta(r-a)]/a $, when
$k_M \rightarrow \infty$ the expectation value $ \langle D(a,r) \rangle_n $
is finite. In terms of this distribution we can define the 
principal value  
\begin{equation}
{\cal P} [ \frac{3}{2} f(0,k_M,r) 
+ \frac{ \delta(r) }{r^2} \ln \frac{k_M}{k_R} ] 
\vert_{k_M \rightarrow \infty } = 
4 \pi \delta^3 (r)  \ln \frac{ \eta }{ a_\mu k_R}
-  {\rm lim}_{a \rightarrow 0 } D(a,r) ~~.
\end{equation}
The choice of a scale factor $\eta=e^{\frac{4}{3} - \gamma}$, where
$\gamma$ is the Euler's constant, yields the formula used by Araki 
\cite{ara}
\begin{equation}
\langle W_{ij} \rangle_n =
-\frac{2 \alpha^3 }{3 \pi } \{ Q_{ij}^n + 4 \pi \langle \delta^3 (r_{ij})
 \rangle_n [ \ln a_\mu k_R - \frac{4}{3}] \}  f^3 ~{\rm a.u.}~,
\end{equation}
where $Q_{ij}^n= {\rm lim}_{a \rightarrow 0} \langle D(a,r) + 4 \pi \gamma
\delta^3(r_{ij}) \rangle_n$. In Ps$^-$ this yields for the effective two-body
contribution $\delta_2 E_n= \delta_{2L} E_n + \sum_{i<j} \langle W^S_{ij}
\rangle_n = \sum_{i<j} \langle W_{ij} \rangle_n$ the expression
\begin{equation}
\delta_2 E_n = -\frac{2 \alpha^3 }{3 \pi } [ Q_{12}^n +Q_{13}^n+Q_{23}^n
\end{equation}
$$
+ 4 \pi ( \ln a_\mu k_R  - \frac{4}{3} )
\langle n \vert \delta^3 (r_1)  + \delta^3(r_2) + \delta^3 (r) \vert n \rangle ] 
f^3 ~{\rm a.u.}~.
$$
Summarizing these results, the effective QED contribution of order $\alpha^3$
to the energy level $E_n$ of Ps$^-$  due to the exchange of a transverse 
photon is  $\delta_{1p} E_n =\delta_1 E_n + \delta_2 E_n$. This sum is
independent of the arbitrary energy unit ${\rm R}_M$, as it should, but for
the purpose of numerical calculations we choose ${\rm R}_M=f {\rm Ry}$.
With this choice, $a_\mu k_R= \alpha /2$, and  $mc/(2 \hbar k_R) =
1/(f \alpha^2)$. \\ \indent
The corresponding terms for positronium can be obtained from the
expressions given above simply by neglecting all the expectation
values containing the variables $r_2$ and $r$, involving the second
electron.  For the Ps ground state $Q_{13}^{Ps}= - 4 \ln 2$, while
 $\beta_g^{Ps}$ is the same as the Bethe logarithm for hydrogen
$\beta_g^H=2.984~128~555~765~497~611(4)$, each Bethe logarithm being
calculated using the corresponding reduced Rydberg constant \cite{dra5}. 
For the Ps$^-$ ground state the numerical values of $Q^g_{12}$ and $Q^g_{13}$ 
used in the present estimates are listed in Table V, while  
$\beta_g = 3.005~030(2)$ \cite{dra5} (including the finite mass correction). 
\\ \indent
To the same order we should consider also the double photon exchange term 
(including the Coulomb part) $\delta_{2p} E_n$ \cite{ara} \cite{pach}
\begin{equation}
\delta_{2p} E_n = -\frac{ \alpha^3 }{2 \pi } [ Q_{12}^n +Q_{13}^n+Q_{23}^n
\end{equation}
$$
- 4 \pi ( \ln f \alpha  - \frac{4}{3} \ln 2 + \frac{13}{6})
\langle n \vert \delta^3 (r_1)  + \delta^3(r_2) + \delta^3 (r) \vert n \rangle ] 
f^3 ~{\rm a.u.}~~,
$$
and the energyy shift associated with the two-photon decay.
\\ \indent
In general, any 
coupling which makes the levels unstable produces a complex energy shift 
$\Delta_c E_n = \delta_c E_n - i \Gamma^n_c /2 $, where $\delta_c E_n$ 
is a correction to the level centroid, $\lambda^n_c = \Gamma^n_c / \hbar$ is 
the decay rate, and $c$ denotes the decay channel. Neutral positronium 
normally decays by spontaneous 
$e^+e^-$ annihilation into two photons if the total spin $S_{ep}=0$, 
($\vec{S}_{ep}= \vec{s}_e + \vec{s}_p$), and in three photons if 
$S_{ep}=1$ \cite{foot}.
The corresponding decay rates are such that $\Gamma_{3 \gamma} \sim
\alpha \Gamma_{ 2 \gamma}$, and the first correction arises from 
two photon annihilation. In this channel $\delta_{ 2 \gamma}  E_n /
\Gamma^n_{ 2 \gamma}  = - (1- \ln 2)/ \pi $ \cite{rkar}, where  
\begin{equation}
\Gamma^n_{ 2 \gamma} =  2 \pi \alpha^3 \langle
( 2 -  \vec{S}^2_{ep}) \delta^3 (r_1 ) \rangle_n f^3 ~ {\rm  a.u.}~. 
\end{equation}
For the Ps ground state ($\vec{S}_{ep}=0$) this yields a decay rate 
$\lambda_{(Ps, 2 \gamma)}  = \alpha^3 {\rm Ry} / \hbar = 8.04 
{\rm nsec}^{-1}$,  in good agreement with the experimental result 
$7.99(11) {\rm nsec}^{-1}$ \cite{ ther}.   \\ \indent
In the Ps$^-$ ground state the electron spins are coupled to 0, and 
the two-photon annihilation can take place between the positron and
any of the two electrons.  The total rate depends on  
$\langle \vec{S}_{13}^2 + \vec{S}_{23}^2 \rangle_g = 3$, and can be
expressed in the form 
\begin{equation}
\Gamma_{(Ps^-, 2 \gamma)}  = 
 2 \pi \alpha^3  \langle \delta^3 (r_1) \rangle f^3 ~ {\rm  a.u.}~. 
\end{equation}
The effective ground state expectation value 
$\langle \delta(  \vec{R}_1 /a_0 )  \rangle $ given in the Appendix, 
Table IV,  yields $\lambda_{(Ps^-, 2 \gamma) }   = 
2.092~797(1)$  nsec$^{-1}$, in good agreement with 
the previous estimates  \cite{bhat} and the experimental result 2.09$(9)$
nsec$^{-1}$ \cite{mill}. The corresponding level shift is
$\delta_{ 2 \gamma} E_g  = - (1- \ln 2) \Gamma_{(Ps^-, 2 \gamma)} / \pi$.
\\ \indent
Summarizing the results of these calculations, the effective ground state
expectation values of the nonrelativistic Hamiltonian and the first
relativistic and QED corrections for Ps and Ps$^-$ are collected in the
following table \\

\noindent
\( 
\begin{array}{lccc} 
E [ {\rm Ry}] & {\rm Ps} & {\rm Ps}^- & E_{\rm Ps}- E_{\rm Ps^-}                \\
      E_g    &  -1/2 & -0.524~010~140~465~960~215~38 
& 0.024~010~140~465~960~215~38  \\
         &       & \pm 0.56 \times 10^{-18} & \pm 0.56 \times 10^{-18}  \\
\langle H_1 \rangle / \alpha^2   &  - 5/32  &  -0.161~254~673~938~50(6)  &
 0.005~004~673~938~50(6)     \\
\langle H_2 \rangle / \alpha^2   &   1/4    & 0.259~466~645~837(8)  & 
-0.009~466~645~837(8)   \\
\langle M_2 \rangle / \alpha^2   &  -1/4   &  -0.243~688~156~296~0(1)  &  
-0.006~311~843~704~0(1)   \\
\langle H_{vp} \rangle / \alpha^3 &  - 1/( 15 \pi )  &   
-0.022~024~212~934~6(7)  & 0.000~803~553~855~7(7)  \\
\delta_{1p} E_g / \alpha^3 &  2.766~873~00(3) &  3.006~491~9(9)   & 
                                                     -0.239~618~9(9)  \\
\delta_{2p} E_g / \alpha^3  &  -0.585~335~778(7)  & -0.510~831~605(7)  & 
-0.074~504~17(1)   \\
\delta_{ 2 \gamma} E_g / \alpha^3 &  -(1- \ln 2 )/ \pi  &
 -0.025~448~161~055(1) &   -0.072~226~124~976(1)        \\

\end{array} \)                     
  \\

In Ps$^-$ the nonrelativistic one electron binding energy 
0.024 010 140 465 960 215 38$(56)$ Ry is practically 
the same as the one determined in \cite{frol}, and close to the older 
estimate of $0.024~010~113 $ Ry \cite{bhat}. The effect of the correction 
terms is to decrease slightly this energy to
\begin{equation}
B'=[0.024~010~140~465~960~215~38(56)  -0.010~773~815~602(8) \alpha^2
\end{equation}
$$
-0.385~545~7(9) \alpha^3 ] ~{\rm Ry} =
0.024~009~416~924~85(6) ~ {\rm Ry}~.
$$
In positronium the observed hyperfine splitting of
$1.160~963(9) \alpha^2$ Ry between the otherwise degenerate ground 
state components corresponds to an additional spin-spin contact term  
\cite{rkar}
\begin{equation}
\delta_s^{ep} E_n^0  = 
2 \pi \alpha^2 \langle \delta^3 (r_1) \{
\frac{4}{3} \vec{s}_e \cdot \vec{s}_p ( 1 - \frac{ \alpha}{2 \pi} )
+ \frac{1}{2} \vec{S}_{ep}^2 [ 1- ( \frac{26}{9} + \ln 4) \frac{
\alpha}{ \pi} ] \} \rangle_n~f^3~{\rm a.u.}~,
\end{equation}
shifting the energy of the singlet by  $ \delta_s^{ep} E_g^0 = 
- 2 \alpha^2 (1 - \alpha /2 \pi) f^3 {\rm a.u.} = 
-0.265~947~576(23) \times 10^{-4} {\rm Ry} $.  
In Ps$^-$ the electron-electron spin-spin dependent energy shift is \cite{ara}
\begin{equation}
\delta_s^{ee} E_g  = - \frac{8 \pi}{3} \alpha^2 (1+ \frac{5}{2 \pi}
 \alpha ) \langle \vec{s}_1 \cdot \vec{s}_2   \delta^3(r) \rangle f^3~{\rm a.u.}~,
\end{equation}
while the two electron-positron spin-spin contact terms contribute by
\begin{equation}
\delta_s^{2ep} E_g = 
\pi \alpha^2 \langle \delta^3 (r_1) 
( \vec{S}_{13}^2 + \vec{S}_{23}^2 ) [ 1- ( \frac{26}{9} + \ln 4) \frac{
\alpha}{ \pi} ]  \rangle ~f^3~{\rm a.u.}~~. 
\end{equation}
These formulas yield an additional shift of the Ps$^-$ ground state 
energy $\delta_sE_g= \delta_s^{ee}E_g + \delta_s^{2ep} E_g =
 0.207~196~744(18) \times 10^{-4}$ Ry, and a contribution to its
binding energy of $ \delta_s^{ep} E_g^0 - \delta_sE_g = 
 -0.473~144~32(3) \times 10^{-4} {\rm Ry} $.
Including the spin-spin contact terms, the binding energy becomes
\begin{equation} 
B''= B' + \delta_s^{ep} E_g^0 -\delta_sE_g =
0.023~962~102~492(3) {\rm Ry}~~.
\end{equation}
\\[.5cm]
{\bf IV. Summary and Conclusions }
\\[.5cm]  \indent
The calculation of the relativistic and QED corrections to the energy 
levels of a quantum three-body system represents a challenging problem 
the modern theory. Difficulties appear both at conceptual and
computational levels, as there is no satisfactory relativistic many-body 
quantum theory, and the nonrelativistic problem is not integrable. 
 \\ \indent
A quantum three-body system thoroughly investigated since the early days of
quantum mechanics is the helium atom. In this system a major simplification
occurrs, because the reduced electron mass $\mu$ is smaller than the mass 
of the positive charge by a factor $1.3707 \times 10^{-4}$, and to a first
approximation the relative motion of the nucleus in the center of mass frame
can be neglected. The relativistic invariance is partly restored by the Breit
interaction, and highly accurate nonrelativistic wave functions can be
obtained numerically, from variational calculations. Within this framework, 
a perturbative treatment of the relativistic and QED corrections gives 
energy levels in remarkable agreement with experiment \cite{dra2}
 \cite{dra3}. \\ \indent
The same procedure was applied in this work to the negative positronium ion.
However, by contrast to helium, all three particles have equall mass,
and a perturbative treatment of the positron motion becomes
inappropriate.  \\ \indent
The accuracy of the nonrelativistic energy and variational ground 
state wave function was discussed in Sect. II. The effective value of 
$E_g$ estimated here is $-0.262~005~070~232~980~107~69(28)$ a.u., the same as
in \cite{frol}, \cite{dra4} and close
within $10^{-8}$ to estimates obtained by other methods \cite{kriv}. 
The variance of the  Hamiltonian for the largest (2528-dimensional) basis set is 
$2.78 \times 10^{-8}$ a.u., smaller than the level  width $\Gamma = \hbar 
\lambda= 5.06 \times 10^{-8}$ a.u. due to the $(2 \gamma) e^-$ decay.
\\ \indent
The calculation of the first relativistic and QED corrections has been 
presented in Sect. III. Formal expressions of these correction terms 
have been known for a long time, but the mass polarization term, which 
cannot be neglected for Ps$^-$, increases dramatically the complexity of the 
numerical calculations. In this work the calculations have been performed
by representing the action of the physical operators on the Hylleraas basis 
states in terms of a suitable set of elementary operators (Appendix, Table I).
The numerical values obtained for some of the most important matrix elements
are summarized in the Tables II-V of the Appendix. It was found that the 
spin independent relativistic terms contribute to the Ps$^-$ ground state 
energy by  $ -0.072~738~092~198(4)  \alpha^2 $ a.u.
and the lowest order QED corrections by $1.224~094~00(44) \alpha^3 $ a.u..
These terms decrease the ground
state energy to $E_{g^*}^{eff}=-0.262~008~467~959~9(4) $ a.u..
Both contributions decrease slightly also the one electron binding energy,
from the nonrelativistic value $0.012~005~070~232~980~10(3)$ a.u. to 
$0.012~004~708~462~43(3) $ a.u.. 
A much larger contribution appears however from the spin-dependent contact 
terms, which raise the ground state energy to 
$E_{g^*}^{eff}= -0.261~998~108~122(1) $ a.u., and
further decrease the binding energy to $0.011~981~051~246(2) $ a.u.. The
calculated decay rate by two photon emission is 2.092 797$(1)$ nsec$^{-1}$,
close to the previous theoretical results and to the measured value
\cite{mill}.
\\ \indent
The binding and ground state energies are shifted also by relativistic
corrections of order $\alpha^3$Ry containing spin-orbit 
magnetic interactions. These contributions are also important, and should 
be included in the further attempts to improve the accuracy of the present 
estimates. \\[1cm]

{\bf Appendix} \\

{\it (a) Elementary operators}  \\

\indent 
In terms of the elementary operators defined in Table I, the 
nonrelativistic Hamiltonian contains linear combinations such as 
$$
(- \frac{1}{2} { \nabla}_1^2 - \frac{1}{2} { \nabla}_2^2) \Phi =
(E_2 +E_3-E_4+E_5 +E_6-0.5 E_9 
$$
$$ -0.5 E_{10} -0.5 E_{11} -0.5E_{12}) \Phi $$
for the kinetic energy part, $\tilde{V} \Phi= - E_{13}\Phi$ for the potential
energy and
$$
{ \nabla}_1 \cdot { \nabla}_2 \Phi= (E_1+E_2+E_3-E_4+E_5+E_6 +E_7+E_8) \Phi
$$
for the recoil term. The action on the basis states of a linear combination 
$\hat{O}_p= \sum_k f_{pk} E_k $, 
has the general form
$$
\hat{ O}_p \Phi_{x_0} = \sum_{t=1}^{N_p} q_{px}^t { A}_t \Phi_{x_t} 
$$
where $x$ denotes the whole set of indices of $\Phi$, $q_{px}^t$ is a factor 
determined by $x$ and the action of $E_k$ on the radial functions, while 
${ A}_t$ denotes the remaining angular operator. In this sum the same 
state ${ A}_t \Phi_{x_t}$  (or ${ A}_t { A}_{t'}  \Phi_{x_{tt'}}$ in
the case of a product) may appear several times with different scalar 
factors $q$. Therefore, in
general it is possible to reduce the number of terms from $N_p$ to 
$N_r<N_p$ by partial summations, before the effective calculation of the
matrix elements. The reduction increases the speed of the numerical 
calculation,  because $N_r$ can be significantly smaller than $N_p$. 
For example, $N_p:N_r$ is $20:18$ in the case of 
$\vec{ \nabla}_1 \cdot \vec{ \nabla}_2$,
$400:235$ for $( \vec{ \nabla}_1 \cdot \vec{ \nabla}_2 )^2$ , while for
$H_0$ and $H_0^2$  is $31:24$  and $961:427$, respectively. \\ \indent
In the computer program the action of
$E_k$ consists of both arithmetic and symbolic operations. The 
arithmetic operations are determined by its action on the radial 
function $r_1^a r_2^b r_{12}^c e^{-\alpha r_1 - \beta r_2}$. 
The number $n_k$ of distinct polynomial terms generated by this action 
is given in the second column of Table I. 
The symbolic operation corresponds to the action of  $E_k$ on the 
spherical harmonics in $ {\cal Y}^{l_1 l_2}_{LM} $, and is coded by a 
character denoting angular operators such as $\hat{r}_1 \cdot 
\hat{r}_2$ or $\nabla^Y_1 \cdot \nabla^Y_2$. \\

{\it (b) Error estimates} \\

\indent
The series of numerical values presented in Tables II-V appear to
be convergent, but for comparison with experiment, it is useful to provide
also a single effective value, representing the expected result of
the present calculation when $N_b \rightarrow \infty$.
The procedure adopted here to define this value depends on the manner of
convergence.  In the case of a sequence $\{ f_n \}$
convergent as an alternating series, the effective value $f_{eff} \pm
\sigma_f $, given in the last row, was defined as the arithmetic average of
its last three consecutive terms, by $ f_{eff} = ( f_{n_x}+f_{n_y}+
f_{n_z} )/3$, $n_x < n_y < n_z $, and $\sigma_f^2 = [ (f_{n_x}-f_{eff})^2+
(f_{n_y}-f_{eff})^2 + (f_{n_z}- f_{eff})^2]/3$. If $\{ f_n \}$ approaches the
limit by monotonous increase or decrease, then we can assume that the series
can be extended to infinity by the function $F(n) = f_{eff}+Ae^{- \gamma n}$.
The matching equations $F(n_x)=f_{n_x}$,
$F(n_y)=f_{n_y}$, $F(n_z)=f_{n_z}$ between $F(n)$ and the last three
calculated numerical
values yield the parameter $f_{eff}$ in the form \cite{dra4}
$$
f_{eff} = f_{n_y} + \frac{ f_{n_y} -f_{n_x} }{R-1}~~.
$$
Here $R \equiv e^{\gamma(n_y-n_x)}$ is the solution of the equation
$R-1=R_y [1-R^{(n_y-n_z)/(n_y-n_x)} ] $, where 
$$
R_y=\frac{ f_{n_y} - f_{n_x}}{f_{n_z} - f_{n_y}} ~~.
$$
The error is assumed to be
$$\sigma_f = \vert f_{n_y} - f_{eff} \vert =   \vert
\frac{ f_{n_y} -f_{n_x} }{R-1} \vert ~~.
$$
If $n_y-n_x =  n_z-n_y$, then $R=R_y$. When $n$ is simply $N_b$, then
$n_y-n_x=286$ is larger, but close to $n_z-n_y=252$, and  $R = R_y$ still
provides a reasonable estimate. \\

{\it (c) Expectation values of $p^4$ } \\ 

\indent
The expectation values $\langle \nabla_1^4 \rangle = \langle
\nabla_2^4 \rangle $ for electrons in
the Ps$^-$ ground state can be calculated numerically either directly, as
$$
\langle \nabla_1^4 \rangle =4 \langle (E_2-0.5E_4+E_6-0.5E_9-0.5E_{11})^2 \rangle
$$
or as $\langle \nabla^4_1 \rangle_E= 0.5 \langle \nabla_1^4+ \nabla_2^4
\rangle_E =0.5 \langle ( \nabla_1^2+ \nabla_2^2)^2 \rangle_E
- \langle \nabla_1^2 \nabla_2^2 \rangle $,  where the first term is expressed in the
form
$$
0.5 \langle ( \nabla_1^2+ \nabla_2^2)^2 \rangle_E = 
2 \langle ( \tilde{E}_g - \tilde{ V} + f { \nabla}_1 \cdot { \nabla}_2)^2 \rangle
$$
$$ = 2 [ \tilde{E}^2_g - 2 \tilde{E}_g \langle \tilde{V} - f { \nabla}_1 \cdot { \nabla}_2
\rangle +
\langle ( \tilde{V} - f { \nabla}_1 \cdot { \nabla}_2)^2 \rangle ]
$$
($\tilde{E}_g \equiv E_g/f{\rm a.u.}$) by assuming that the variational
ground state is practically eigenstate of $H_0$. Although formally the same
at the limit $N_b \rightarrow \infty$, the numerical values obtained for
$\langle \nabla^4_1 \rangle$ and $\langle \nabla^4_1 \rangle_E$ at finite
$N_b$ are slightly different.
These estimates are given as a function of the basis size $N_b$ in the 
first two columns of Table III. The third column contains the relativistic 
correction term for the positron, given by
$$
\langle \nabla_3^4 \rangle_E = \langle ( { \nabla}_1+ { \nabla}_2)^4
\rangle_E = 2 \langle \nabla^4_1 \rangle_E +
2 \langle \nabla_1^2 \nabla_2^2 \rangle - 8 \langle ( \tilde{E}_g - \tilde{V} )
{ \nabla}_1 \cdot { \nabla}_2 \rangle~~.
$$
\vskip.3cm
{\it (d) QED corrections of order $\alpha^3$ 
in the limit $m_3 \rightarrow \infty$ } \\

\indent 
When $m_3 \rightarrow \infty$ the vacuum polarization term and the 
contribution of the electron-electron spin dependent contact interaction  
$\delta_s^{ee} E_n \vert_{ \alpha^3}= -(20/3) \langle \vec{s}_1 \cdot \vec{s}_2 
\delta^3(r) \rangle_n \alpha^3 f^3$ a.u. remain the same, but 
$X^n_3=0$, $Y^n_{i3}=0$, and $\delta_{1p}E_n$ becomes
$$
 \delta_{1p} E_n^\infty = \alpha^3 [
  \frac{ 4 }{ 3 } ( \ln \frac{mc}{2 \hbar k_R} + \frac{5}{6} 
-  \beta_n ) \langle \delta^3(r_1) + \delta^3(r_2) \rangle_n 
$$
$$
- \frac{ 8}{3}  ( \ln \frac{mc}{2 \hbar k_R} - \frac{1}{2}
+ \ln a_\mu k_R) \langle \delta^3(r) \rangle_n
- \frac{2}{3 \pi} Q_{12}^n ]f^3~{\rm a.u.}~.
$$
The two-photon contribution reduces to
$$
\delta_{2p} E_n^\infty = \alpha^3 [ - \frac{Q_{12}^n}{2 \pi} +
2  ( \ln f \alpha - \frac{4}{3} \ln 2 + \frac{13}{6} ) 
\langle \delta^3(r) \rangle_n ] f^3~{\rm a.u.}~~,
$$
and the total correction 
$\delta E_n^\infty = \delta_{1p} E_n^\infty + \delta_{2p} E_n^\infty
+ \delta_s^{ee} E_n \vert_{\alpha^3} + \langle H_{vp} \rangle_n$   is
$$
\delta E_n^\infty = \alpha^3 [ \frac{4}{3} ( \frac{19}{30} -
 \ln f \alpha^2 - 
 \beta_n ) \langle \delta^3(r_1) + \delta^3(r_2) \rangle_n 
+ ( \frac{14}{3} \ln f \alpha + \frac{164}{15})  \langle \delta^3(r) 
\rangle_n
- \frac{7}{6 \pi} Q_{12}^n ]f^3~{\rm a.u.}~~.
$$ 
\vskip1cm
{\bf Table captions} \\
Table I. Elementary operators $\{ E_k, k=1,15 \}$ used in the calculation
of the expectation values. $n_k$ denotes the number of distinct polynomial
terms generated by the action of $E_k$ on a basis state. \\
Table II. The ground state expectation values $E_g= \langle H_0 \rangle$,
$\langle H_0^2 \rangle $ and $\sigma^2 = \langle H_0^2 \rangle
-E_g^2$ as a function of the basis dimension $N_b$. \\
Table III. Ground state expectation values of the singular
differential operators $\langle \nabla^4_1 \rangle_E$,
$\langle \nabla^4_1 \rangle $ and $\langle \nabla^4_3 \rangle_E$
as a function of the basis dimension $N_b$. \\
Table IV. Ground state expectation values of the singular Dirac distributions
$\langle \delta ( \vec{R}_1 ) \rangle$ and $\langle \delta( \vec{R} ) \rangle
$ as a function of the basis dimension $N_b$. \\
Table V. The ground state expectation values 
$Q^g_{12}$ and $Q^g_{13}$ defined by
$Q_{ij}^g = {\rm lim}_{a \rightarrow 0} \langle \theta (r_{ij}-a)
/r^3_{ij} + 4 \pi [ \gamma + \ln (a/a_\mu) ] \delta^3 ( r_{ij} ) \rangle
$, as a function of the basis dimension $N_b$. \\

\newpage

\begin{center}

{\bf Table I}

\vskip.5cm
\(
\begin{array}{rcl} 
k & n_k & E_k \\
1 & 4 & \hat{r}_1 \cdot \hat{r}_2  \partial^2_{r_1 r_2 } \\
2 & 4 & r^{-1} (r_2  \hat{r}_1 \cdot \hat{r}_2   -r_1) \partial^2_{r_1 r} \\
3 & 4 & r^{-1} (r_1  \hat{r}_1 \cdot \hat{r}_2   -r_2) \partial^2_{r_2 r} \\
4 & 1 & \partial^2_r+2 r^{-1} \partial_r \\
5 & 1 & r_1( r_2r)^{-1} \hat{r}_1 \cdot \nabla_2^Y \partial_r \\
6 & 1 & r_2 (r_1r)^{-1} \hat{r}_2 \cdot \nabla_1^Y \partial_r \\
7 & 4 & r_2^{-1} \hat{r}_1 \cdot \nabla_2^Y \partial_{r_1} + r_1^{-1} 
 \hat{r}_2 \cdot \nabla_1^Y \partial_{r_2} \\
8 & 1 & (r_1r_2)^{-1}  \nabla_1^Y \cdot \nabla_2^Y  \\
9 & 3 & \partial^2_{r_1} + 2 r_1^{-1} \partial_{r_1} \\
10 & 3 & \partial^2_{r_2} +2 r_2^{-1} \partial_{r_2}  \\
11 & 1 &  r_1^{-2} \nabla_1^Y \cdot \nabla_1^Y  \\ 
12 & 1 &  r_2^{-2} \nabla_2^Y \cdot \nabla_2^Y  \\ 
13 & 3 & r_1^{-1}+ r_2^{-1}- r^{-1} \\
14 & 1 & -l_1(l_1+1) r_1^{-2} \\
15 & 1 & -l_2(l_2+1) r_2^{-2} \\
\end{array} \)

\end{center}
\vskip.5cm

\newpage

\begin{center}
\vspace{.5cm}
{\bf Table II} 
\vspace{.5cm}

\(
\begin{array}{rccc} 
 N_b & E_g [{\rm Ry}] & \langle H_0^2 \rangle [{\rm Ry}^2] &  \sigma^2 \times
 10^{14} [{\rm Ry}^2] \\
324 & -0.524~010~140~413~399~000~28  &  0.274~586~632~449~596  & 519~352.588~780~183~9 \\
411 & -0.524~010~140~455~551~566~88  &  0.274~586~628~565~868  & 126~562.113~711~645~6 \\
512 & -0.524~010~140~464~139~040~54  &  0.274~586~627~626~769  & 31~752.227~055~472~07 \\
630 & -0.524~010~140~465~665~621~87  &  0.274~586~627~375~932  & 6~508.536~436~039~192 \\
764 & -0.524~010~140~465~918~375~12  &  0.274~586~627~323~102  & 1~199.051~083~460~867 \\
918 & -0.524~010~140~465~954~391~13  &  0.274~586~627~313~704  & 255.472~732~778~647~0 \\
1089 & -0.524~010~140~465~959~038~66  &  0.274~586~627~311~885 & 73.086~762~209~023~06 \\
1283 & -0.524~010~140~465~960~002~45  &  0.274~586~627~311~421 & 26.589~554~852~763~07 \\
1495 & -0.524~010~140~465~960~160~85  &  0.274~586~627~311~222 & 6.672~153~792~304~994 \\
1733 & -0.524~010~140~465~960~203~19  &  0.274~586~627~311~175 & 1.965~916~264~831~472 \\
1990 & -0.524~010~140~465~960~212~96  &  0.274~586~627~311~165 & 0.965~692~558~621~058 \\
2276 & -0.524~010~140~465~960~214~82  &  0.274~586~627~311~160 & 0.465~897~417~244~748 \\
2528 & -0.524~010~140~465~960~215~25  &  0.274~586~627~311~158 & 0.308~152~352~372~668 \\
 eff & -0.524~010~140~465~960~215~4(6)  & 0.274~586~627~311~156(4)  & 0.23(23) \\
\end{array} \)
\end{center}

\begin{center}

\vspace{.5cm}
{\bf Table III}
\vspace{.5cm}

\(
\begin{array}{rccc} 
 N_b & \langle \nabla^4_1 \rangle_E &  \langle \nabla^4_1 \rangle  &
 \langle \nabla^4_3 \rangle_E \\
  324 &  2.532~451~004~442~6  & 2.532~445~719~29 & 5.255~396~862~891 \\
  411 &  2.532~451~050~420~6  & 2.532~451~697~56 & 5.255~397~122~254 \\
  512 &  2.532~451~056~877~0  & 2.532~450~741~84 & 5.255~397~117~353 \\
  630 &  2.532~451~009~132~0  & 2.532~449~964~61 & 5.255~397~051~034 \\
  764 &  2.532~451~018~719~1  & 2.532~450~992~21 & 5.255~397~086~467 \\
  918 &  2.532~451~022~453~6  & 2.532~451~056~52 & 5.255~397~094~127 \\
 1089 & 2.532~451~021~529~7 & 2.532~451~023~43 & 5.255~397~091~672 \\
 1283 & 2.532~451~020~589~3 & 2.532~451~019~49 & 5.255~397~091~024 \\
 1495 & 2.532~451~020~595~0 & 2.532~451~022~17 & 5.255~397~090~993 \\
 1733 & 2.532~451~020~587~2 & 2.532~451~020~24 & 5.255~397~090~958 \\
 1990 & 2.532~451~020~559~2 & 2.532~451~019~92 & 5.255~397~090~940 \\
 2276 & 2.532~451~020~559~6 & 2.532~451~020~43 & 5.255~397~090~949 \\
 2528 & 2.532~451~020~560~0 & 2.532~451~020~42 & 5.255~397~090~945 \\
eff  & 2.532~451~020~559~6(3) & 2.532~451~020~2(2) & 5.255~397~090~945(4) \\
\end{array} \)
\end{center}

\newpage

\begin{center}
\vspace{.5cm}
{\bf Table IV}
\vspace{.5cm}

\( 
\begin{array}{rcc} 
 N_b &   \langle \delta ( \vec{R}_1) \rangle [a_0^{-3}] & \langle
 \delta ( \vec{R} ) \rangle [a_0^{-3}] \\
 324 & 0.020~733~174~230~2 & 0.000~171~000~000~8  \\
 411 & 0.020~733~203~838~1 & 0.000~170~999~383~2  \\
 512 & 0.020~733~199~804~5 & 0.000~170~999~967~2  \\
 630 & 0.020~733~193~292~2 & 0.000~170~997~306~7  \\
 764 & 0.020~733~197~986~7 & 0.000~170~996~885~4   \\
 918 & 0.020~733~198~238~9 & 0.000~170~996~811~0   \\
 1089 & 0.020~733~198~094~3 & 0.000~170~996~832~4  \\
 1283 & 0.020~733~197~999~5 & 0.000~170~996~756~0  \\
 1495 & 0.020~733~198~024~3 & 0.000~170~996~767~3  \\
 1733 & 0.020~733~198~007~4 & 0.000~170~996~760~1  \\
 1990 & 0.020~733~198~003~4 & 0.000~170~996~757~7  \\
 2276 & 0.020~733~198~005~3 & 0.000~170~996~757~1  \\
 2528 & 0.020~733~198~005~0 & 0.000~170~996~756~8  \\
eff & 0.020~733~198~004~6(8) & 0.000~170~996~756~7(4)    \\
\end{array} \)                                                        
\end{center}

\begin{center}
\vspace{.5cm}
{\bf Table V}
\vspace{.5cm}

\( 
\begin{array}{rcc} 
 N_b &     Q_{12}^g [a_\mu^{-3}]      &   Q_{13}^g [a_\mu^{-3}]  \\
324  & 0.095~757~780~75 & -2.776~563~295 \\
411  & 0.095~757~975~79 & -2.776~588~343  \\
512  & 0.095~757~804~27 & -2.776~583~829  \\
630  & 0.095~758~749~78 & -2.776~578~687 \\
764  & 0.095~758~904~03 & -2.776~582~810 \\
918  & 0.095~758~930~40 & -2.776~582~894 \\
1089 & 0.095~758~918~40 & -2.776~582~776 \\
1283 & 0.095~758~949~78 & -2.776~582~694 \\
1495 & 0.095~758~944~76 & -2.776~582~722 \\
1733 & 0.095~758~947~86 & -2.776~582~703 \\
1990 & 0.095~758~949~04 & -2.776~582~700 \\
2276 & 0.095~758~949~31 & -2.776~582~702 \\
eff  & 0.095~758~949~4(3) & -2.776~582~702(1)  \\
\end{array} \)                                                        
\end{center}

\end{document}